\documentclass[aps,prb,preprint,superscriptaddress]{revtex4-1}
\usepackage[breaklinks=true,colorlinks,citecolor=blue,linkcolor=blue,urlcolor=blue]{hyperref}

\usepackage{amssymb,amsmath,amsfonts}
\usepackage{subfigure, latexsym,multirow}
\usepackage[notcite,color,final]{showkeys}
\usepackage{epsfig,graphicx,subfigure,dcolumn,bm}
\usepackage[usenames ,dvipsnames]{xcolor}
\usepackage{slashed}
\usepackage{ulem}
\DeclareMathOperator\sign{sgn}

\begin{document}
\title{Transverse circular photogalvanic effect  associated with  Lorentz-violating Weyl fermions}

\author{Mohammad Yahyavi$^*$\footnote[0]{*These authors contributed equally to this work.}}
\affiliation {Division of Physics and Applied Physics, School of Physical and Mathematical Sciences, Nanyang Technological University, Singapore, Singapore}

\author{Yuanjun Jin$^*$}
\affiliation {Division of Physics and Applied Physics, School of Physical and Mathematical Sciences, Nanyang Technological University, Singapore, Singapore}

\author{Yilin Zhao}
\affiliation {Division of Physics and Applied Physics, School of Physical and Mathematical Sciences, Nanyang Technological University, Singapore, Singapore}

\author{Zi-Jia Cheng}
\affiliation {Laboratory for Topological Quantum Matter and Advanced Spectroscopy (B7), Department of Physics, Princeton University, Princeton, New Jersey 08544, USA}

\author{Tyler A. Cochran}
\affiliation {Laboratory for Topological Quantum Matter and Advanced Spectroscopy (B7), Department of Physics, Princeton University, Princeton, New Jersey 08544, USA}

\author{Yi-Chun Hung}
\affiliation {Institute of Physics, Academia Sinica, Taipei 115229, Taiwan}

\author{Tay-Rong Chang}
\affiliation {Department of Physics, National Cheng Kung University, Tainan, Taiwan}
\affiliation {Center for Quantum Frontiers of Research and Technology (QFort), Tainan, Taiwan}
\affiliation {Physics Division, National Center for Theoretical Sciences, Taipei 10617, Taiwan}

\author{Qiong Ma}
\affiliation {Department of Physics, Boston College, Chestnut Hill, Massachusetts 02467, USA}
\affiliation {CIFAR Azrieli Global Scholars program, CIFAR, Toronto, Canada}

\author{Su-Yang Xu}
\affiliation {Department of Chemistry and Chemical Biology, Harvard University, Cambridge, MA, USA}

\author{Arun Bansil}
\affiliation{Department of Physics, Northeastern University, Boston, MA 02115, USA}

 \author{M.~Zahid~Hasan}\affiliation {Laboratory for Topological Quantum Matter and Advanced Spectroscopy (B7), Department of Physics, Princeton University, Princeton, New Jersey 08544, USA}
\affiliation{Lawrence Berkeley National Laboratory, Berkeley, California 94720, USA}

\author{Guoqing Chang  $^{\dag}$\footnote[0]{$^{\dag}$ Corresponding author (email): guoqing.chang@ntu.edu.sg }}
\affiliation{Division of Physics and Applied Physics, School of Physical and Mathematical Sciences, Nanyang Technological University, Singapore, Singapore}

\date{\today}

\maketitle
\clearpage
 \newpage

\textbf{Nonlinear optical responses of quantum materials have recently undergone dramatic developments to unveil nontrivial geometry and topology. A remarkable example is the quantized longitudinal circular photogalvanic effect  (CPGE) associated with the Chern number of Weyl fermions, while the physics of transverse CPGE in Weyl semimetals remains exclusive. Here, we show that the transverse CPGE of Lorentz invariant Weyl fermions is forced to be zero. We find that the transverse photocurrents of Weyl fermions are associated not only with the Chern numbers but also with the degree of Lorentz-symmetry breaking in condensed matter systems.  Based on the generic two-band model analysis, we provide a new powerful equation to calculate the transverse CPGE based on the tilting and warping terms of Weyl fermions. Our results are more capable in designing large transverse CPGE of Weyl semimetals in experiments and are applied to more than tens of Weyl materials to estimate their photocurrents. Our method paves the way to study the CPGE of massless or massive quasiparticles to design next-generation quantum optoelectronics.
}\\

The past few years have witnessed rapid progress in our understanding of nonlinear optical responses of topological quantum materials~\cite{Hasan2010TI,Zhang2011,Bansil2016,Armitage2018,Nagaosa2020Transport,Hasan2021Weyl,Ma2021Topology,Orenstein2021,Lv2021Exp,Bao2022,acoustics2022}.  Nonlinear electromagnetics creates new opportunities to uncover, probe, and control wavefunction geometry and topology which offers promise for future applications~\cite{mciver2012control,Wu2016,chan2017photocurrents,Ma2017,Wu2017Giant,xu2018Electrically,BCYin2022,BCMa2022}.  In non-magnetic materials, circular photogalvanic effects (CPGE) inducing a non-oscillating direct current can directly probe the band-resolved Berry curvature~\cite{Ma2017,Nagaosa2020Transport,Orenstein2021}.  In this regard, Weyl semimetals hosting magnetic monopoles in momentum space  provide an ideal platform to measure Berry curvature~\cite{Armitage2018,xu2015,LVPRX2015}. It has been demonstrated theoretically that longitudinal CPGE of a chiral Weyl semimetal is quantized to the Chern number of Weyl fermions \cite{de2017quantized1,chang2017unconventional,Chang2018,Flicker2018}. While recent experiments realized and verified the signature of quantized CPGE in topological chiral crystal RhSi family~\cite{RhSiExp1,RhSiExp2,RhSiExp3,RhSiExp4,RhSiExp6,rees2020helicity,ni2020linear,ni2021giant,RhSiExp5}. Surprisingly, the case of transverse CPGE is more important in practical application because the current flows perpendicular to the direction of the polarization of incident light which can be easily detected and utilized in next-generation light sensors, solar cells, and information processing. Despite intensive experimental studies and the giant transverse photocurrents detected in different Weyl materials \cite{Yan2017,Chang2018,Ma2019Nonlinear,Osterhoudt2019,liu2020semimetals,Ma2021Topology,acoustics2022,mciver2012control,Yuan2014TCPGE,Ma2017,chang2020unconventional}, a comprehensive and fundamental physical interpretation of the transverse CPGE in Weyl semimetal is still exclusive. This hinders the progress of the application of Weyl semimetals as the next-generation quantum photodetectors.

Therefore, it is crucial to explore the physics behind transverse CPGE and reveal the criteria to observe a large value of it. 
 The second-order tensor of  injection current is written as \cite{de2017quantized1,chang2017unconventional}
\begin{equation}\label{CPGE_Org}
	\beta_{ij}(\omega)=\frac{\pi e^3}{\hbar^2}\epsilon_{j\beta\gamma} \int_{\textbf{k}} \sum_{n,m} f^{\textbf{k}}_{nm} \Delta^i_{\textbf{k},mn} r^\beta_{\textbf{k},nm} r^\gamma_{\textbf{k},mn} \delta(\omega_{mn}-\omega)
\end{equation}
where $\int_{\textbf{k}}=\int  d^d k/(2\pi)^d$, $\hbar \omega_{nm}=\hbar \omega_{n}-\hbar \omega_{m}$ is the energy difference between bands $n$ and $m$, $f_{n}$ is the Fermi-Dirac distribution of the band $n$, $f_{nm}=f_{n}-f_{m}$, $\Delta_{mn}^i=\partial_{k_i}E_{mn}/\hbar$, and $r_{nm}^\beta =i\langle  u^n_{\vec{k}}|\partial_{k_\beta}| u^m_{\vec{k}}\rangle$ is the off-diagonal Berry connection or interband transition matrix element. Numerically, one of the obstacles is too expensive to achieve transverse CPGE based on the product of a matrix element and joint density of state using Eq. \ref{CPGE_Org}~\cite{de2017quantized1} because it requires a very high density of $k$-mesh near the Weyl nodes to obtain reliable results.

\begin{figure}[t!] 
	\includegraphics[width=\columnwidth]{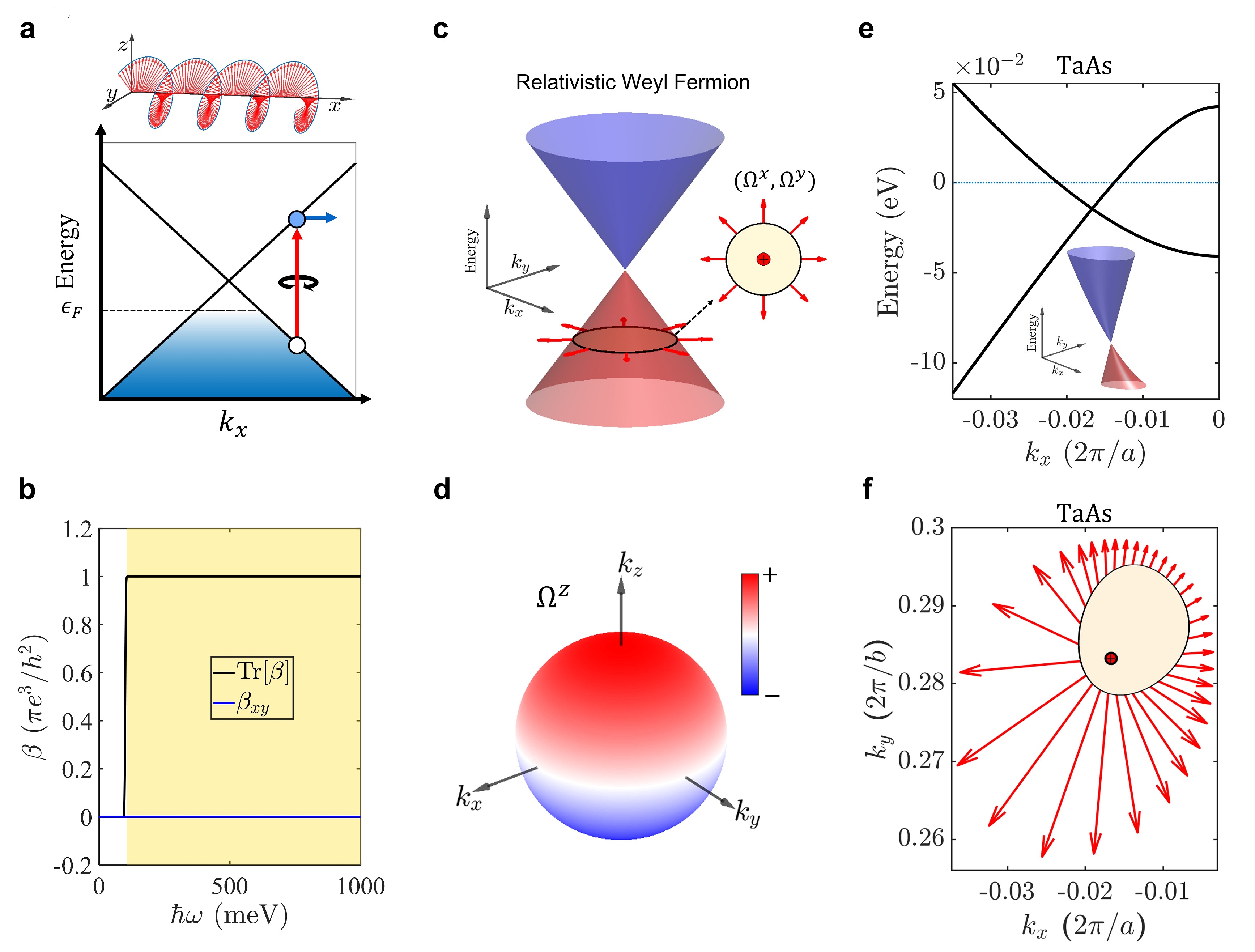}
	\caption{{\textbf{CPGE from isotropic Weyl fermion.} \textbf{(a)} Top: Schematic of circularly polarized laser. Bottom: Schematic of the CPGE from a relativistic Weyl fermion. An electron (filled blue circle) is excited from conduction bands to the valance band by circularly polarized laser (red arrow), inducing injection current (blue arrow). Where states below Fermi level ($\epsilon_{F}=-50$ meV) are shaded blue to indicate electron filling. \textbf{(b)} Trace and off-diagonal components of injection current $\beta$  calculated from the isotropic relativistic Weyl fermion illustrated in panel \textbf{a}. All the off-diagonal components of the CPGE tensor are to be zero, while the trace of $\beta$ is quantized.  \textbf{(c)} Left: Three-dimensional band structure of an isotropic Weyl fermion. Right: In-plane Berry curvature components  ($\Omega^x$,$\Omega^y$) at the black Fermi contour.	The red arrows show the flow of the Berry curvature field. \textbf{(d)}  Distribution of Berry curvature component $\Omega^z$ on a relativistic Weyl fermion's Fermi surface.  $\Omega^z$ is odd along $z$-direction, but even along $x$ and $y$-directions. \textbf{(e)} Band structure of a Weyl fermion in TaAs material. In the real material, the Weyl cones will have certain tilting and warping terms in Hamiltonian to deform the shape of Weyl cones. Inset figure shows three-dimensional band structure for fixed  $k_z$ momenta.  \textbf{(f)} A Fermi contour and Berry curvature distribution  ($\Omega^x$,$\Omega^y$)  on it  is asymmetric about the Weyl point. 
	}} 
	\label{Fig.1}
\end{figure}

To study the CPGE response of a Weyl semimetal, we first start with a relativistic Weyl Hamiltonian:
\begin{equation}
	\mathcal{H}(\textbf{k})=  \sum_{i=x,y,z} \nu_i \hbar k_i \sigma_i 
	\label{Eq.SWH1}
\end{equation}
where $\nu_i$ is velocity, $\hbar$ is Planck's reduced constant, $k_i$  is momentum and  $\sigma_i$ are the Pauli matrices. We can see that, for isotropic relativistic Weyl fermions (Fig. \ref{Fig.1}a), the CPGE trace is quantized which are consistent with previous theory  by Tr$[\beta(\omega)]=i\mathcal{C} \pi e^3/\hbar^2$, where $\mathcal{C}=+ 1$ is the monopole charge of the Weyl fermion (Fig. \ref{Fig.1}b)~\cite{de2017quantized1}. In contrast, all the off-diagonal components of the CPGE tensor for isotropic relativistic Weyl fermions are zero. 

To better understand the result, we perform symmetry analysis concerning the Weyl point in the momentum space. In a relativistic Weyl Hamiltonian, an injection current of ${\beta}_{ij}(\omega)$ is proportional to the integral of $\tilde{\beta}_{ij}(\textbf{k},\omega)\propto k_i \Omega^j(\textbf{k})$ at the \textbf{k}-points where optical transition happened, here $\Omega^j(\textbf{k})$ is Berry curvature component along $j$-direction. For a relativistic Weyl fermion which preserves the Lorentz symmetry, its dispersion and Berry curvatures are isotropic along all directions (Figs. \ref{Fig.1}c and \ref{Fig.1}d). In this case, concerning the Weyl point in the momentum space,  $\Omega^j(\textbf{k})$ is an odd function in $j$-direction and even in other directions (details are in Supplementary Section I).  Thus, in this condition,  $\tilde{\beta}_{ii}(\textbf{k},\omega)$ ($\tilde{\beta}_{ij}(\textbf{k},\omega)$) is an even (odd) with respect to $\textbf{k}$. 
This explicitly demonstrates that  the integral of $\tilde{\beta}_{ij}(\textbf{k},\omega)$ over the Fermi surface is zero, indicating the off-diagonal elements of CPGE vanishes for the relativistic Weyl node.

Isotropic Weyl semimetal is equivalent to the Lorentz invariant Weyl fermion and respects the Lorentz symmetry \cite{soluyanov2015type,LVPRX2015} (Fig. \ref{Fig.1}c).  In fact, the actual band dispersion in the real materials will inevitably have warping and tilting, which break the Lorentz invariance of Weyl fermions from quantum field theory. In Fig. \ref{Fig.1}e, we show the band structure of TaAs around a Weyl node~\cite{xu2015,huang2015weyl,WangPRX2015}, showing the Weyl node is not strongly isotropic and does not respect the Lorentz symmetry. In this case, the Berry curvature penetrating through a closed Fermi surface is asymmetric, so anisotropic excitation can occur (Fig. \ref{Fig.1}f). 

Our above analysis shows that the Lorentz violation terms guarantee the off-diagonal components of the CPGE tensor. Next, we aim to establish exact correspondence between the off-diagonal components of the CPGE tensor and the Lorentz violation term by inviting new formulae and comparing the results with that of Eq. \ref{Eq.SWH1}. We next discuss under what conditions the off-diagonal components of the CPGE tensor remain non-zero. 

 \textbf{The effects of warping corrections.}
 
\begin{figure}[t!] 
	\includegraphics[width=1\columnwidth]{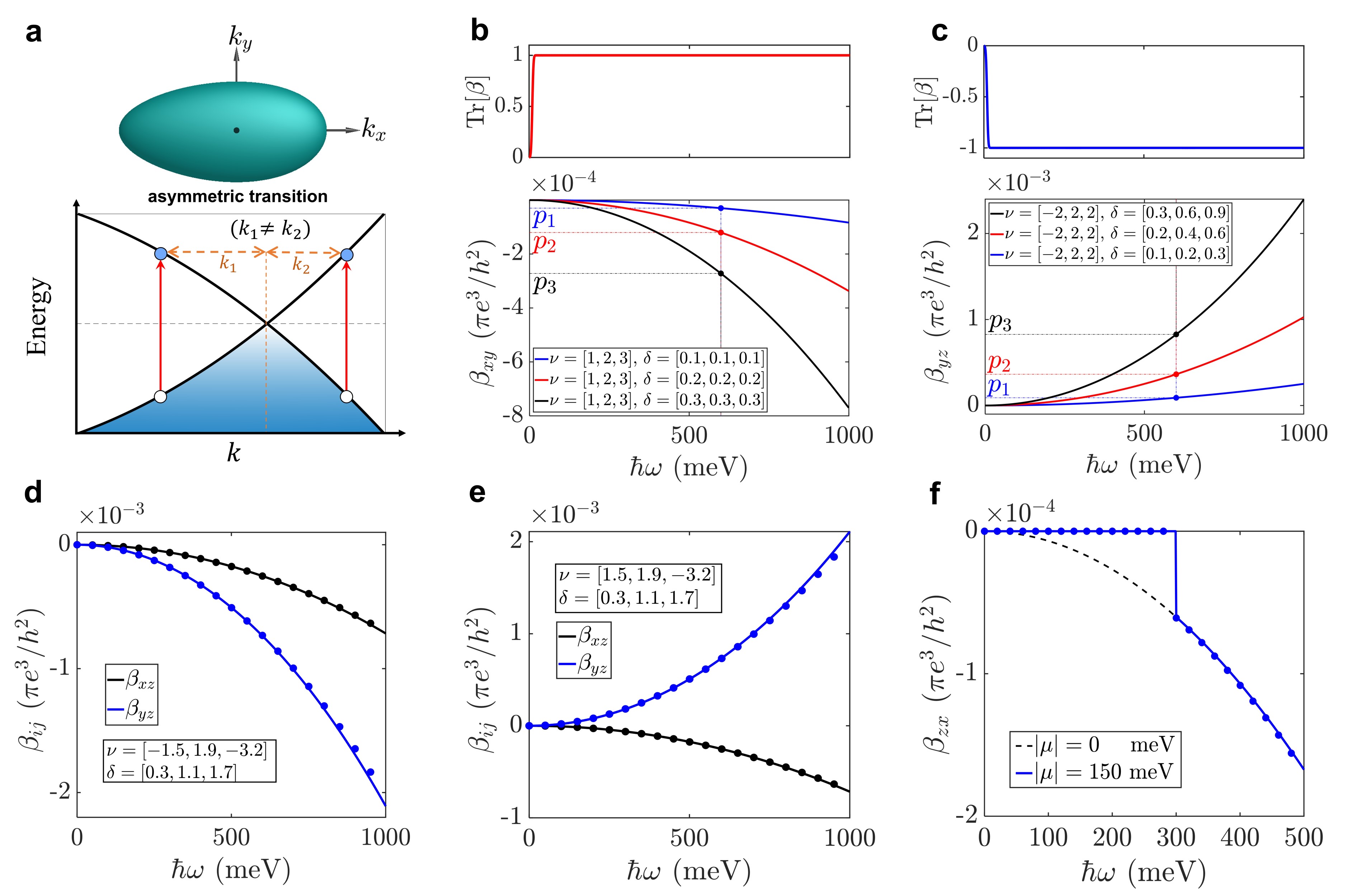}
	\caption{\textbf{Transverse CPGE of Weyl fermion due to quadratic corrections.} \textbf{(a)} Top: The Fermi surface of Weyl fermion with the warping corrections term. In this case, the Fermi surface is an asymmetric tapered oval centered away from the nodal point, so anisotropic excitation can occur. Bottom: Schematic of Weyl fermion with the warping correction  terms.   Transitions of electron (filled blue circle)  from conduction bands to the valance band by circularly polarized laser (red arrow) for two parts of the spectrum with the same photon energy is asymmetric in the region of momenta $k$ ($k_1 \neq k_2$).  \textbf{(b)} Top: Trace of CPGE of Weyl node with positive chirality which is quantized in units of $ \pi e^3/h^2$. Bottom: Transverse CPGE of $\beta_{xy}$ for same condition of top in panel \textbf{b}.   Here $p_1=-2.997\times 10^{-5}$, $p_2=-1.202\times 10^{-4}$, and $p_3=-2.720\times 10^{-4}$. \textbf{(c)} Top: Trace of CPGE of Weyl node with negative chirality  which is quantized in units of $- \pi e^3/h^2$. Bottom: Transverse CPGE of $\beta_{yz}$ for same condition of top in panel \textbf{c}. Here $p_1=9.009\times 10^{-5}$, $p_2=3.629\times 10^{-4}$, and $p_3=8.266\times 10^{-4}$. Off-diagonal components of the CPGE tensor from our proposed formula of Eq. \ref{BB} (dotted line) and directly numerical calculation from the Eq. \ref{CPGE_Org} (solid line) for \textbf{(d)}  $\nu=[-1.5,1.9,-3.2]$, and \textbf{(e)}  $\nu=[1.5,1.9,-3.2]$ with fixed value of $\delta=[0.3,1.1,1.7]$.}
	\label{Fig.2}
\end{figure}

Firstly, we consider the warping correction  terms in the Weyl Hamiltonian expressed as
\begin{equation}
	\mathcal{H}(\textbf{k})=   \sum_{i=x,y,z}  \left(\nu_i \hbar k_i+\delta_i \hbar^2 k_i^2 \right)   \sigma_i 
	\label{Eq.GWH}
\end{equation}  
where $\nu_i$ is velocity, $\delta_i$ is warping corrections to Hamiltonian, and $\sigma_i$ are the Pauli matrices. In the limit, $\delta_i= 0$, the model reduces to the Hamiltonian describing a standard Weyl point. After the inclusion of warping corrections, the Fermi surface is an asymmetric tapered oval centered away from the nodal point (Fig. \ref{Fig.2}a, top panel). As a result, for a fixed frequency of the laser, the optical transitions becomes anisotropic around the Weyl fermions (Fig. \ref{Fig.2}a, bottom panel).

To make a canonical formalism and precise that relates transverse CPGE with the warping correction terms of Hamiltonian, we first consider the case of the same value of $\delta_i$ in all three directions (Fig. \ref{Fig.2}b). Consistent with our previous analysis, nonzero transverse CPGE appears due to the breaking of Lorentz symmetry. Using $\beta_{xy}$ as an example, one can see it scales linearly with the power of two of the warping term $\beta_{xy} \propto\delta^2$. Same for all the other transverse CPGE components (details are in Supplementary Section  II).  Then, we consider another parameter set, where the velocity terms are changed to reverse the Chern number and the warping terms are different (Fig. \ref{Fig.2}c). We can see the scaling between the transverse CPGE component and the warping terms is still valid. What is more, compared to Fig. \ref{Fig.2}b, the signs of transverse CPGE components are flipped in Fig. \ref{Fig.2}c. This indicates transverse CPGE component of a Weyl fermion is also related to its Chern number. We can also find that the amplitude of the transverse CPGE component decreases with the increase of velocity terms of Weyl fermions. Having determined the detailed transverse CPGE of different models (details are in Supplementary Section II), we can elucidate the correspondence between the warping correction and the transverse CPGE response. Our detailed analysis shows that the off-diagonal components of the CPGE tensor of the warping correction  terms in the Weyl Hamiltonian have the form 
\begin{equation}
	\beta_{ij} =  \frac{-2e^3 \sign(\mathcal{C})}{3\pi h^2} \left[ \frac{\delta_i  \delta_j }{\nu_i\nu_j^3}    (\hbar \omega)^2 \right] \label{Bij00}
\end{equation}   
The fitting performance of the above equation to the numerically direct calculation of the off-diagonal components of the CPGE tensor are compared in Fig. \ref{Fig.2}d,e for some arbitrary values of $\nu_i$ and $\delta_i$. It can be seen that there is good agreement between both spectra. 
 
In Eq. \ref{Bij00}, we are setting the Fermi level exactly at the energy of Weyl nodes. In real materials, there will be an energy difference $\mu$ between the  Weyl point energy and the Fermi energy ($\mu=\epsilon_w-\epsilon_F$). To further examine the effect of $\mu$ on photocurrent, we compute the transverse CPGE as a function of photon energy for different values of  $\mu$ (Fig. \ref{Fig.2}f). The abruptness of the jump in photocurrents at the characteristic photon energy,  $2|\mu|$,  depends on the Fermi energy and can be expressed as:
\begin{equation}\label{BB}
	\beta_{ij} =  \frac{-2e^3 \sign(\mathcal{C})}{3\pi h^2} \left[ \frac{\delta_i  \delta_j }{\nu_i\nu_j^3}    (\hbar \omega)^2 \;\Theta(\hbar \omega-2\mu)\right] 
\end{equation}    
where  $\Theta(x)$ is Heaviside step function.  Comparing the fitting performance of the above equation with the numerically direct calculation shows that this formula corresponds to a clear physical picture of transverse CPGE of Weyl Hamiltonian with the warping correction terms.  Our   Eq. \ref{BB} shows that obtaining a large transverse CPGE response requires band structures to possess strong warping terms.


\textbf{The effects of tilted velocity.} 

Another way to generate the off-diagonal components of the CPGE tensor is to consider the velocity tilting of the cone term in the Hamiltonian. The most general low-energy effective model  describing a  Weyl Hamiltonian has this form 
\begin{equation}
	\mathcal{H}(\textbf{k})=   \sum_{i=x,y,z}   \left(\nu_i^t \hbar k_i+ \nu_i \hbar k_i    \sigma_i \right) -\mu
	\label{Eq.GWH}
\end{equation}
where $\nu_i$ is velocity, $\nu^t_i$ is tilting velocity which breaks the degeneracy between the bands at $\textbf{k}$ and $\textbf{-k}$,  $\epsilon_F$ is chemical potential, $\sigma_i$ are the Pauli matrices.   

Different from warping terms which deform the shape of Fermi surfaces, titling terms will only shift relative position between a Weyl fermion and its exited Fermi surface under a laser, while the Fermi surface and its Berry curvature distribution remain isotropic (details are in Supplementary Section II). 
However, anisotropic optical excitation can still happen in Weyl fermions with tilting under a laser of low frequency.  
As shown in Fig. \ref{Fig.3}a, when the energy of the photon is larger than $ \epsilon_{1}$ but smaller than $ \epsilon_{2}$, only electrons on the one side of the Weyl cone will be excited while the optical transition another side is forbidden.  It corresponds to transitions for which only part of the spectrum is Pauli blocked.  As a result, a hole will appear on the excited Fermi surface under a fix frequency of laser, inducing anisotropic optical excitation within the photon energy interval $ \epsilon_{1}<\hbar \omega<\epsilon_{2}$ (Fig. \ref{Fig.3}b). The energy interval is determined by the direction of the largest titling: $\epsilon_{1,2}=\frac{2|\mu|}{1\pm\mathcal{W}^T}$, where $\mathcal{W_T} =  \sqrt{(\frac{\nu_x^t}{\nu_x})^2+(\frac{\nu_y^t}{\nu_y})^2+(\frac{\nu_z^t}{\nu_z})^2}$ is the dimensionless tilt parameter which reveals the largest degree of tilting of the Weyl cone.  (details are in Supplementary Section II).

In order to make a canonical formalism and the connection precise that relates transverse CPGE with the tilted terms of Hamiltonians, we followed the same steps above. Our analysis shows that the off-diagonal components of the CPGE tensor of the tilted term in the Weyl Hamiltonian have the form
\begin{equation}
	\beta_{ij} =   \frac{3e^3\sign(\mathcal{C}) }{\pi h^2} \left[\frac{\nu_i^t \nu_j^t}{\nu_j^2}   \frac{1}{\mathcal{W_T}^2}    \sum_{n=1}^2 (-1)^n \sin(\gamma_n)\cos(\gamma_n)^3 \right]\label{Eq.tB}
\end{equation}
with
\begin{equation}
	\gamma_n = \arcsin\{\frac{1}{\mathcal{W_T}}[\frac{2|\mu|}{\hbar\omega}+(-1)^{n+1}]\}
\end{equation}

To confirm the correctness of our formula, we investigated off-diagonal components of the CPGE tensor for arbitrary values of $\nu_i$ and $\nu^t_i$ in Fig. \ref{Fig.3}c (details are in Supplementary Section II). First, we can see, indeed, in the energy window ($\epsilon_{1}<\hbar \omega<\epsilon_{2}$) where anisotropic optical excitation happens, there is a nonzero transverse CPGE. This is both captured by calculation based on original Eq. \ref{CPGE_Org} (Fig. \ref{Fig.3}c, solid blue curve) and our formula Eq. \ref{Eq.tB} (Fig. \ref{Fig.3}c, dotted curve). Second, inside the nonzero transverse CPGE allowed window,  our results based on formula Eq. \ref{Eq.tB} match well with the direction calculations. 


 Along some specific directions in \textbf{k}-space, if the two dispersions forming the Weyl cone have the same sign of Fermi velocity, it is called the type-II Weyl nodes (Fig. \ref{Fig.3}d). In type-II Weyl semimetals, for any value of the $\mu$ when the energy of a photon is larger than $ \epsilon_{1}$, only electrons on the one side of the Weyl cone will be excited while the optical transition on other side is forbidden (Fig. \ref{Fig.3}d,e). Surprisingly, the equation \ref{Eq.tB} we get from type-I Weyl  fermions ($\mathcal{W_T}<1$) is still valid for  type-II Weyl fermions \cite{soluyanov2015type} ($\mathcal{W_T}>1$).  To examine this, the fitting performance of the Eq. \ref{Eq.tB}  to the results of an exact numerical calculation of the transverse  CPGE  using the Eq. \ref{CPGE_Org} are compared in Fig. \ref{Fig.3}f for some arbitrary values of $\nu$ and $\nu^t$. It can be seen that there is good agreement between both spectra.

Our formula can also help us find the condition where the titling terms can give rise to large transverse CPGE photocurrent analytically. Specifically, we focus on the peaks of the CPGE from Eq. \ref{Eq.tB}(details are in Supplementary Section II). We find that $\beta_{ij}$ is maximized at the condiction $\nu_k^t/\nu_k=0$ and  $\nu_i^t/\nu_i=\nu_j^t/\nu_j$.  In an attempt to confirm this finding, we calculate the phase diagram as a function of the $\nu_i^t/\nu_i$  and the $\nu_j^t/\nu_j$ with $\nu_k^t/\nu_k=0$.  To obtain the maximum value of transverse CPGE, the peak value is checked within the photon energy in the case anisotropic excitation can occur. By calculating the normalized $\beta_{ij}$ based on Eq. \ref{Eq.tB}, we can see the maximum value of off-diagonal CPGE is significantly affected by tuning tilting velocity in $i$ and $j$-direction (Fig. \ref{Fig.3}g). Now we verify our finding via the original formula Eq. \ref{CPGE_Org}. We can see, consistent with our analysis based on Eq. \ref{Eq.tB}, $\beta_{xy}$ is maxed when $\nu_z^t/\nu_z=0$ and  $\nu_x^t/\nu_x=\nu_y^t/\nu_y$ for both type-I (Fig. \ref{Fig.3}h) and type-II (Fig. \ref{Fig.3}i) Weyl fermions.  For the type-I Weyl fermions, the peak will decrease quickly when changing the parameters away from  $\nu_z^t/\nu_z=0$ and  $\nu_x^t/\nu_x=\nu_y^t/\nu_y$ (Fig. \ref{Fig.3}h). While for type-II Weyl fermions, the peak will change slowly when change the titling parameters. 
   
Now we combine warping terms and tilting terms in Weyl equation,
\begin{equation}
	\mathcal{H}(\textbf{k})=  \sum_{i=x,y,z} \left(\nu_i^t \hbar k_i + \left(\nu_i \hbar k_i+\delta_i \hbar^2 k_i^2 \right) \sigma_i \right)  -\mu 
	\label{Eq.GG1}
\end{equation}	
{Then, the generalized transverse CPGE of Weyl fermions has the following form,}
\begin{equation}
	\beta_{ij} = \frac{e^3\sign(\mathcal{C}) }{\pi h^2} \left[\frac{3\nu_i^t \nu_j^t}{\nu_j^2}   \frac{1}{\mathcal{W_T}^2}    \sum_{n=1}^2 (-1)^n \sin(\gamma_n)\cos(\gamma_n)^3 - \frac{2\delta_i  \delta_j }{3\nu_i\nu_j^3}    (\hbar \omega)^2 \;\Theta(\hbar \omega-2\mu)\right]
	\label{Eq.GG2}
\end{equation}

The first part of our Eq. \ref{Eq.GG2} from the titling terms has a dominant contribution to the transverse CPGE current tensor in the low frequency but the second part from the warping terms has a dominant contribution in the high frequency.  We can also see that the transverse CPGE from the tilting term of a  Weyl node must vanish exactly at $\hbar \omega =2|\mu|$ (details are in Supplementary Section  II). The transverse CPGE from the warping term when the $\hbar \omega < 2|\mu|$
In other words, if this photon energy is very close but smaller than $2|\mu|$, the measured transverse CPGE from Weyl will be approaching zero. Thus measuring the transverse CPGE of Weyl semimetals also provide a powerful tool to measure the energy of Weyl nodes relative to the Fermi level.

\textbf{Transverse CPGE of the real materials.} 

Now we apply our results to the first experimental discovered Weyl semimetal TaAs.  TaAs is a nodal ring semimetal in the absence of spin-orbit coupling (SOC). After turning on SOC, a gap between conduction and valence bands of around 40 meV was open along the high symmetry lines \cite{xu2015,LVPRX2015, huang2015weyl,WangPRX2015}  (Fig. \ref{Fig.4} a). To avoid the effect of massive nodal rings, here we focus on the low-frequency region. In Weyl semimetal TaAs, there are two classes of Weyl points: eight Weyl points on the $k_z = 0$ plane and sixteen Weyl points away from the $k_z = 0$ plane  (Fig. \ref{Fig.4} b). We only select the class of sixteen Weyl nodes since they are near the Fermi level and at low frequency ($\hbar\omega<$40 meV) only this class has contributed to the transverse CPGE (Fig. \ref{Fig.4}b). 

We use our effective model in Eq. \ref{Eq.GG1}, whose parameters, $\nu^t$, $\nu$, and $\delta$, are fitted to DFT band structures of TaAs around a source Weyl point  (Fig. \ref{Fig.4}c).  After that, the parameter of Hamiltonian in Eq. \ref{Eq.GG1} sets obtained directly from the fitting, one can now calculate the off-diagonal components of the CPGE tensor using our formula in Eq. \ref{Eq.GG2}. We also routinely used  Eq. \ref{CPGE_Org} to calculate the CPGE around the source Weyl fermions and the result converges well under the mesh grind of $200^3$. The results of the formula Eq. \ref{Eq.GG2} first presented in this work are in good agreement with the direct calculation of Eq. \ref{CPGE_Org} (Fig. \ref{Fig.4}d). 
 
Once the transverse CPGE of the source Weyl points is known,  we can use our Eq. \ref{Eq.GG2} to get the CPGE of all the other Weyl points according to our symmetric analysis of transverse CPGE and thus the full CPGE can be reconstructed (see Table S1,S2 in Supplementary for details). In TaAs, the only nonzero CPGE are $\beta_{xy}$ and $\beta_{yx}$. According to Eq. \ref{Eq.GG2}, all the six transverse components of the source Weyl point are nonzero (see Table S1,S2 in Supplementary for details). However, for the Weyl point which is related to the source Weyl point by $M_x$ mirror plane, the parameters  $k_x$, $v_x$, $v_x^t$ and $C$ will gain a minus sign while all the other components are the same. As a result, $yz$ and $zy$ components of CPGE from the source Weyl fermion and its partner of mirror plane $M_x$ will cancel with each other. By considering all the symmetries, using Eq. \ref{Eq.GG2} we can also see the off-diagonal CPGE will be nonzero along $\beta_{xy}$ and $\beta_{yx}$ in TaAs, which are consistent with experimental measurements. We now reconstruct the full off-diagonal components of the CPGE tensor of the TaAs from the source Weyl fermion based on $\beta_{ij}  = 8(\beta_{ij}^{S}-\beta_{ji}^{S})$. The reconstruction to the numerically direct calculation of the off-diagonal components of the CPGE tensor is compared in Fig. \ref{Fig.4}d. It can be seen that there is good agreement between both spectra. Our symmetry analysis shows that the above method can be extended to other space groups by calculating independent Weyl points using our invited formulae if the materials belong to one of these point groups $C_{1}$, $C_{2}$, $C_{s}$, $C_{2v}$, $C_{4}$, $S_{4}$, $C_{4v}$, $C_{3}$, $C_{3v}$, $C_{6}$, and $C_{6v}$. See more details in Table S1,S2 in Supplementary information.
 
  This result shows that the value of the non-zero transverse CPGE relied on the contributions from the transverse CPGE of individual Weyl nodes.  
  We want to address that the formula developed in our work is much more powerful than the original Eq. \ref{CPGE_Org} in studying the CPGE of Weyl fermions. It is expensive and time-consuming to study the CPGE of Weyl semimetals at the low-frequency region, especially for the infrared region where Weyl semimetals are good candidates as photodetectors.   It has been found that based on the product of a matrix element and joint density of state~\cite{de2017quantized1},  the transverse CPGE  tends to converge slowly concerning the size $N^3$ number of $k$-points used for the Brillouin zone integration. For example, in TaAs, it l took weeks (Fig. \ref{Fig.4}e) and it cost a memory of the order of terabytes (Fig. \ref{Fig.4}f) to get reliable results of the CPGE using the original formula. In contrast, using our formula the computing time and memory cost are negligible. With this advantage, thus we apply our method to more than other ten known Weyl materials: TaP, NbP, NbAs, LaAlGe, PrAlGe, CeAlGe, Ta$_3$S$_2$, WP$_2$, MoP$_2$, Co$_3$Sn$_2$S$_2$, SrSi$_2$, HgCr$_2$Se$_4$, CeSbTe, Ta$_2$Se$_8$I, and TaIrTe$_4$ \cite{Hasan2021Weyl}. The details of the transverse CPGE calculation of every Weyl semimetal, that we checked by our algorithm are given in Supplementary  Sections  IV,V.

 

Our work unveils that nonzero transverse CPGE of Weyl fermions can exist because of the Lorentz-symmetry breaking in condensed matter physics. Different from longitudinal CPGE in Weyl semimetal which only depends on the Chern number, we show that transverse CPGE has much richer correspondences with the degree of  Lorentz-violation. We for the first time provide a new simple and powerful formula for the transverse CPGE of Weyl fermions, which can be widely applied to a vast of Weyl materials to find better candidates for photodetectors. Beyond Weyl fermions, it is also possible to apply the method developed to other quasiparticles with nontrivial wavefunction geometry, including but not limited to multi-fold chiral fermions, nodal rings, and massless or massive Dirac fermions in both three-dimensional and two-dimensional materials. Our results will greatly accelerate the process of searching for next-generation topological photodetectors.

\section*{Competing interests}
The authors declare no competing interests.

\section*{Acknowledgments}
This work at Nanyang Technological University was supported by the National Research Foundation, Singapore under its Fellowship Award (NRF-NRFF13-2021-0010) and the Nanyang Technological University startup grant (NTUSUG). Work at Princeton University was supported by the Gordon and Betty Moore Foundation (GBMF4547 and GBMF9461; M.Z.H.). The work at Northeastern University was supported by the Air Force Office of Scientific Research under award number FA9550-20-1-0322 and benefited from the computational resources of Northeastern University's Advanced Scientific Computation Center (ASCC) and the Discovery Cluster. The work at National Cheng Kung University was  supported  by  the  2030  Cross-Generation  Young  Scholars  Program from the National Science and Technology Council (NSTC) in Taiwan (MOST111-2628-M-006-003-MY3), National Cheng Kung University (NCKU), Taiwan, and National Center for Theoretical  Sciences,  Taiwan.  This  research  was  supported,  in  part,  by  Higher  Education Sprout  Project,  Ministry  of  Education  to  the  Headquarters  of  University  Advancement  at NCKU.



\addtocounter{figure}{-1}
\begin{figure} [t!]
	\caption{ Our proposed simple fitting function matches our numerical calculation of off-diagonal components of the CPGE tensor from Eq. \ref{CPGE_Org}. \textbf{(e)} At finite dopings, the abruptness of the jump in the transverse CPGE at the characteristic photon energy,  $2|\mu|$,  depends on the Fermi energy.}
\end{figure}
\begin{figure}[t!] 
	\includegraphics[width=\columnwidth]{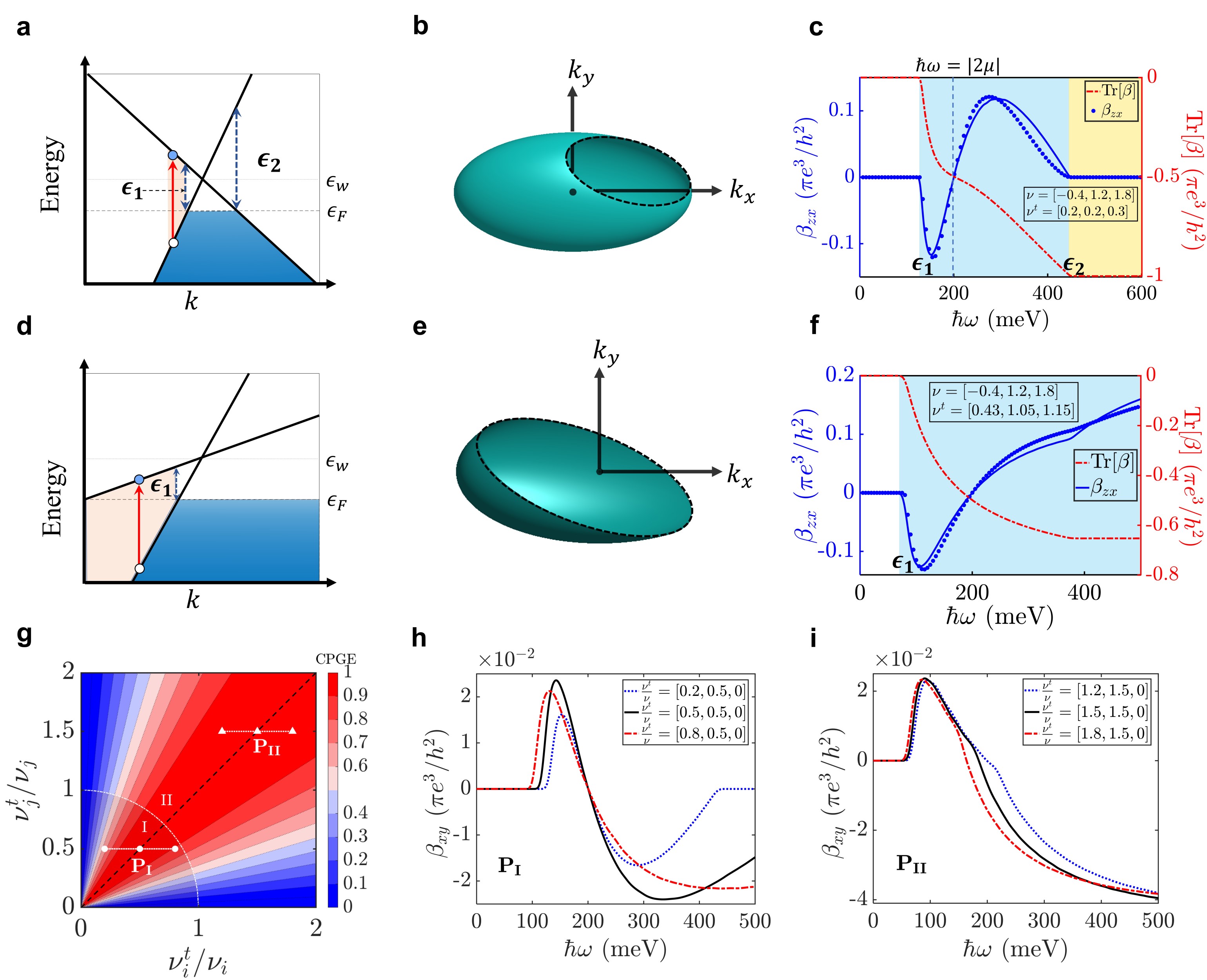}
	\caption{\textbf{Transverse CPGE of Weyl fermion due to the tilted velocity.} \textbf{(a)} Schematic of a two-band system, with degeneracy corresponding to a type-I Weyl fermion. For  $|\epsilon_F|>\epsilon_{\mathcal{W}}$,  the lower and higher edges for the finite size Fermi surface are generated within the photon energy interval $\epsilon_1<\hbar \omega<\epsilon_2$, where $\epsilon_1$ ($\epsilon_2$) gives the lower (higher) edges for the optical absorption. The red arrow indicates representative photon electronic transition (filled blue circle) in the optical absorption region which is highlighted with light coral color. It corresponds to transitions for which only part of the spectrum is Pauli blocked.  \textbf{(b)} Finite size Fermi surface for this photon energy interval ($\epsilon_1<\hbar \omega<\epsilon_2$).  \textbf{(c)} CPGE trace and off-diagonal ones from our proposed formula of Eq. \ref{Eq.tB} (dotted line) and directly numerical calculation from the Eq. \ref{CPGE_Org} (solid line)  of a   type-I Weyl semimetal model for $\mu=-100$ meV chemical potential.  \textbf{(d)} Schematic of a two-band system, with degeneracy corresponding to a type-II Weyl fermion. \textbf{(e)} Finite size Fermi surface for the  the anisotropic excitation energy region ($\hbar \omega>\epsilon_1$). }
	\label{Fig.3}
\end{figure} 
\addtocounter{figure}{-1}
\begin{figure} [t!]
	\caption{\textbf{(f)} CPGE trace and off-diagonal ones of a type-II Weyl semimetal model for $\epsilon_{F}=-100$ meV chemical potential.  Off-diagonal components of the CPGE tensor are to be non-zero, while the anisotropic excitation occurs.  \textbf{(g)} Phase diagram of the gigantic peaks of the transverse CPGE, $\beta_{ij}$, as a function of ${\nu_j^t/\nu_j}$ and ${\nu_i^t/\nu_i}$. The unit is normalized. \textbf{(h)} and \textbf{(i)} Off-diagonal CPGE, $\beta_{xy}$, along two line paths with same length in type-I and type-II cases.}
	\label{Fig.3}
\end{figure}
\clearpage 
\begin{figure}[t!] 
	\includegraphics[width=\columnwidth]{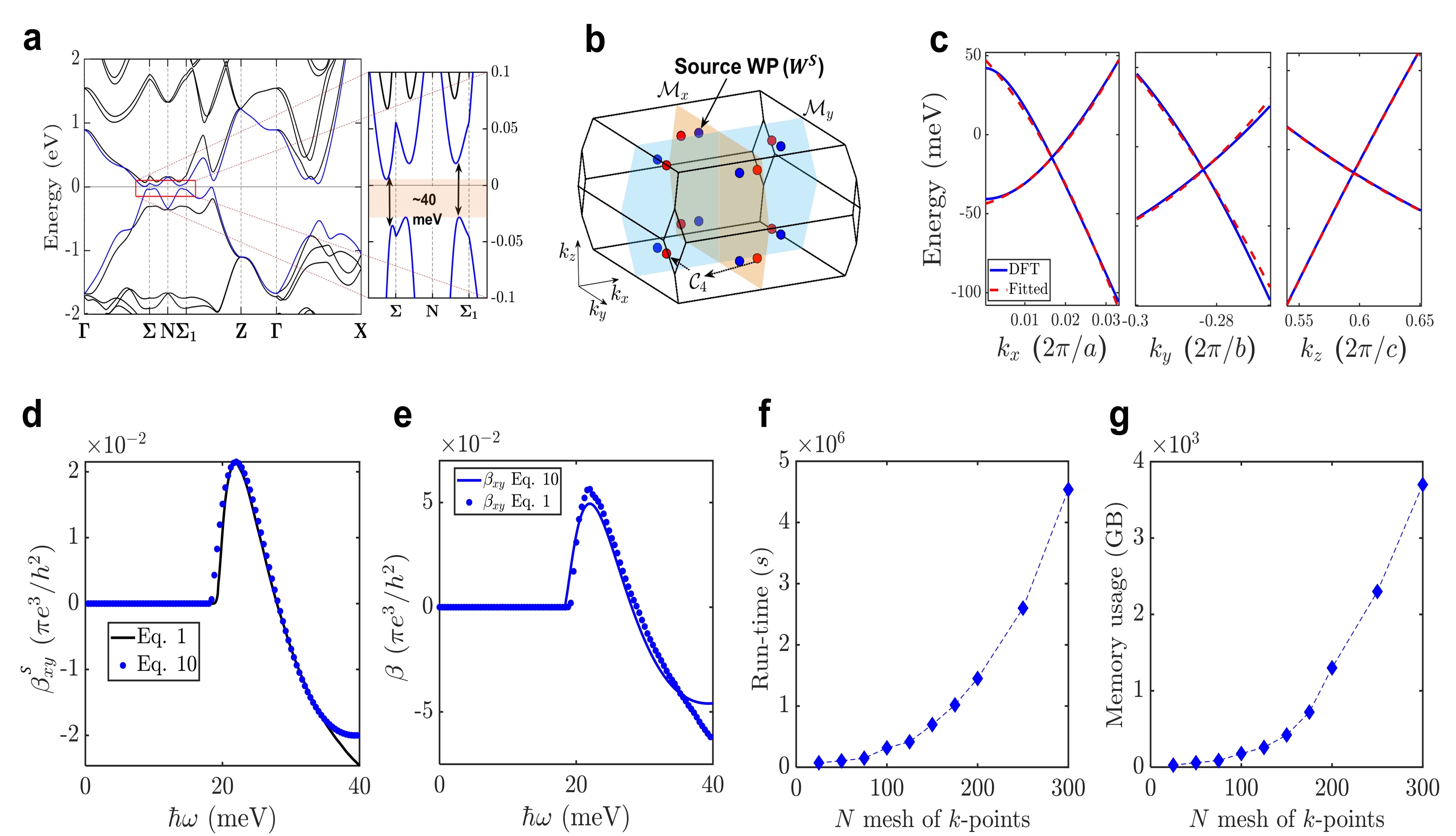}
	\caption{\textbf{Application to Weyl semimetal TaAs.} \textbf{(a)}  Calculated band structure of TaAs in the presence of spin-orbit coupling from DFT calculation. The right panel shows the band structure in the presence of spin-orbit coupling along the $\Sigma$-N-$\Sigma_1$  in
			the vicinity of the Fermi energy. \textbf{(b)} In the presence of spin-orbit coupling, the first Brillouin zone of TaAs material host sixteen  Weyl points denoted as $W_2$ which are related by two mirror plans $\mathcal{M}_x$ and $\mathcal{M}_y$. \textbf{(c)}  Fitted band energy of our model of in Eq. \ref{Eq.GG1} to DFT band structures of TaAs around Weyl point. Where $\nu^t=[-0.5534,   -0.4787,0.9152]$, $\nu=[2.0677,1.7456,-1.8011]$, and $\delta=[18.4201,7.7985,3.0199]$. Off-diagonal components of the CPGE tensor, \textbf{(d)} $\beta_{xy}^S$ from our proposed formula (dotted line) and directly numerical calculation for source Weyl point of  \textbf{(b)}  (solid  line).  $\beta_{yx}^S$ is shown in Supplementary Section V. \textbf{(f)} Directly numerical calculation (solid line) and reconstructed the full off-diagonal components of the CPGE tensor ($\beta_{xy}$) of the TaAs based on source Weyl point data (dotted line).  \textbf{(f)} Run-time and  \textbf{(g)} memory usage for calculating the transverse CPGE spectrum  of the TaAs as a function of the size $N^3$ of integration grid.} 
	\label{Fig.4}
\end{figure} 

\end{document}